\documentclass[aps,preprint]{revtex4}%
\usepackage{amsfonts}
\usepackage{amsmath}
\usepackage{amssymb}
\usepackage{graphicx}
\usepackage{subcaption}%
\begin{document}
\title{Quantum correlation and origin of Hawking radiation for mass-superposed BTZ
black hloes}
\author{Hongbin Zhang}
\author{Baocheng Zhang}
\email{zhangbaocheng@cug.edu.cn}
\affiliation{School of Mathematics and Physics, China University of Geosciences, Wuhan
430074, China}

\begin{abstract}
We investigate the mass superposition of BTZ black holes using the velocity
correlation function. Interestingly, the quantum signatures of BTZ black hole
mass superposition can be revealed by the velocity correlation peaks.
Additionally, different mass superpositions for the same mass ratio can be
distinguished, a phenomenon not previously documented. We also find that the
correlation function method can pinpoint the location where Hawking radiations
are generated, even for mass-superposed BTZ black holes. This supports the
concept of a quantum atmosphere as the origin locus of Hawking radiation, even
in quantum-superposed spacetime.


\end{abstract}
\maketitle

\section{Introduction}

In 1974, Hawking discovered that the black hole can emit thermal radiation
\cite{hawking1974black,hawking1975particle}. The radiation is usually
understood to stem from pairs of correlated particles created near the black
hole horizon, in which one particle escapes to infinity as Hawking radiation,
while the other is swallowed by the singularity inside the black hole horizon,
known as the partner particle. However, the exact location where Hawking
radiation originates remains to be determined, as it can address the
information loss problem
\cite{hawking1976breakdown,almheiri2021entropy,zhang2009hidden,zhang2011entropy}%
. Unruh first discussed the origin of Hawking radiation from gravitational
black holes \cite{unruh1977origin}. Subsequent studies analyzed analog black
holes \cite{schutzhold2008origin} and used two-point correlation functions to
reanalyze gravitational black holes \cite{schutzhold2010quantum}. Later,
Giddings introduced the concept of a quantum atmosphere, displaced from the
horizon, as the origin of Hawking radiation \cite{giddings2016hawking}.
Research on the origin of Hawking radiation has been extended to various
scenarios
\cite{hod2016hawking,dey2017black,buss2018quantum,dey2019black,eune2019test,kaczmarek2024signatures,ong2020quantum}%
, but it has not been investigated for the case in which the spacetime is in
the superpositions.

The concept of spacetime superpositions is not easy to define. A nice way was
given by Bekenstein
\cite{bekenstein1973black,bekenstein2020quantum,hodPhysRevLett.81.4293} that
the black hole should have the discrete eigenvalues for the mass and the
related eigenstates can be in the superpositions. Mass-superposed black holes
are an example of spacetime superpositions; since different masses
individually define a unique classical solution to the Einstein field
equations, and the eigenvalues {}{}of mass superposition correspond to
associated spacetime superposition states. The possibility of black hole mass
superpositions has been explored within the framework of loop quantum gravity
\cite{7,8,9,10,11}. Researchers were also attempting to apply these concepts
to analog black holes \cite{barcelo2022analogue,barcelo2022chronology},
building on earlier work
\cite{unruh1981experimental,barcelo2011analogue,shi2023quantum,zhang2023shielding}%
, and testing quantum effects in gravity through tabletop experiments
\cite{christodoulou2019possibility,marletto2017gravitationally,bose2017spin}.
However, it is not known how to clearly demonstrate the quantum nature of
black hole mass superposition. Until the Ref. \cite{foo2022quantum}, the
quantum signatures of superpositions of mass eigenstates were revealed through
an operational method using the so-called Unruh-DeWitt detector
\cite{crispino2008unruh}. The definition for the mass-superposed black hole or
the black hole mass superpositions have been interpreted above, and more
detailed description can be found in Ref. \cite{foo2022quantum}.

It is well-known that the creation of Hawking radiation is closely related to
the existence of the horizon. When the concept of spacetime superpositions
\cite{zych2019bell,henderson2020quantum,foo2023quantum,suryaatmadja2023signatures}
is introduced, the horizon will be in the superposition. Thus, the origin of
Hawking radiation in this case requires to be investigated again, as implied
by the study of Bekenstein and Mukhanov that mass quantization should alter
the nature of Hawking radiation \cite{gr-qc/9505012}. In this paper, we will
use the correlation function to analyze the occurrence position and evolution
of Hawking radiation for the mass-superposed BTZ black hole, as made in analog
black holes
\cite{balbinot2008nonlocal,carusotto2008numerical,fabbri2021ramp,liao2019proposal,michel2016phonon}%
. The correlation functions have become a key piece of evidence for measuring
Hawking radiation from analog black holes in laboratory settings
\cite{steinhauer2016observation,munoz2019observation}. This method has also
been applied to determine the origin of Hawking radiation from gravitational
black holes
\cite{parentani2010vacuum,balbinot2022quantum,anderson2022horizons,balbinot2023quantum,fontana2023stress,balbinot2023hawking}%
. But whether it is proper to study the origin of Hawking radiation for
mass-superposed BTZ black hole is not clear, and a crucial step is to see
whether it can reveal the quantum signatures of black hole mass
superpositions, as made in Ref. \cite{foo2022quantum}. In this paper, we will
study the correlation functions for mass-superposed BTZ black holes to gain
more insight into mass superposition and the original location of Hawking radiation.

This paper is organized as follows. In the second section, based on the
Wightman functions, we construct the correlation functions for mass-superposed
BTZ black holes, and analyze the role played by the observers. Using the
correlation function outside the event horizon of mass-superposed BTZ black
holes, we demonstrate the quantum signatures of black hole mass superpositions
in the third section. In the fourth section, we investigate the origin of
Hawking radiation using the constructed correlation function across the
horizon of both classical and mass-superposed BTZ black holes. Finally, we
present our conclusions in the fifth section.

\section{Correlation functions}

In (2+1)-dimensions spacetime, Ba\~{n}ados, Teitelboim, and Zanelli found a
solution of the Einstein gravitational field equation called the BTZ black
hole \cite{banados1992black}, whose line element is
\begin{equation}
ds^{2}=-(\frac{r^{2}}{\ell^{2}}-M)dt^{2}+\frac{1}{(\frac{r^{2}}{\ell^{2}}%
-M)}dr^{2}+r^{2}d\phi^{2}, \label{btzm}%
\end{equation}
where $M$ is the mass of the BTZ black hole, $\ell$ $(>0)$ is the anti-de
Sitter (AdS) length scale in an asymptotically AdS vacuum, $\Lambda
=-{1/\ell^{2}}$ is the cosmological constant, and the event horizon is located
at $r_{h}=\sqrt{M}\ell$. It is noted that the BTZ line element given here does
not include the angular momentum, which is convenient to discuss the
mass-superposed black holes. Generally, the BTZ line element can be obtained
from AdS$_{3}$-Rindler spacetime ($\tilde{t},\tilde{r},\tilde{\phi}$) by
rescaling the angular $\tilde{\phi}=\phi\sqrt{M}$, radial coordinate
$\tilde{r}=r/\sqrt{M}$, and time coordinate $\tilde{t}=t\sqrt{M}$. Therefore,
the local geometry related to this BTZ line element is isomorphic to the AdS spacetime.

In the following, we will utilize Wightman functions to get the required
correlation functions. One objective of studying these correlation functions
is to gather additional insights into BTZ black holes in mass superpositions.
Another aim is to pinpoint where the signal indicating the presence of Hawking
radiation emerges in both the classical BTZ black hole and the mass-superposed
BTZ black hole.

Since the BTZ spacetime is a quotient of AdS$_{3}$ spacetime
\cite{banados1992black,bhtz1993}, the Wightman function (also called the
two-point correlation function) of the BTZ spacetime can be obtained by a sum
of correlation functions in AdS$_{3}$ spacetime
\cite{lifschytz1994scalar,carlip19952+},
\begin{equation}
W_{BTZ}^{(1,1)}\left(  x,x^{\prime}\right)  =\frac{1}{\mathcal{N}}\sum_{p}%
\sum_{q}\eta^{p}\eta^{q}\frac{1}{4\pi\ell\sqrt{2}}\left(  \frac{1}%
{\sqrt{\sigma\left(  {\Gamma}_{1}^{p}x,{\Gamma}_{1}^{q}x^{\prime}\right)  }%
}-\frac{\zeta}{\sqrt{\sigma\left(  {\Gamma}_{1}^{p}x,{\Gamma}_{1}^{q}%
x^{\prime}\right)  +2}}\right)  , \label{wf11}%
\end{equation}
which can be obtained from vacuum $AdS$ Wightman function by $W_{BTZ}%
^{(1,1)}\left(  x,x^{\prime}\right)  =\frac{1}{\mathcal{N}}\sum_{p}\sum
_{q}\eta^{p}\eta^{q}\langle0|\hat{\psi}\left(  \Gamma_{1}^{p}x\right)
\hat{\psi}\left(  \Gamma_{1}^{q}x^{\prime}\right)  |0\rangle$, and the vacuum
AdS Wightman function is given as \cite{bhtz1993,smith2014looking}
$W_{AdS}(x,x^{\prime})=\langle0|\hat{\psi}(x)\hat{\psi}(x^{\prime}%
)|0\rangle=\frac{1}{4\pi\ell\sqrt{2}}\left(  \frac{1}{\sqrt{\sigma
(x,x^{\prime})}}-\frac{\zeta}{\sqrt{\sigma(x,x^{\prime})+2}}\right)  $, where
$|0\rangle$ represents the AdS vacuum state, and the parameter $\zeta$ is the
boundary condition at infinity in AdS spacetime, taking the values
$\{0,1,-1\}$ corresponding to transparent, Dirichlet, and Neumann boundary
conditions, respectively. Moreover, $x$ represents $t_{1}$ and $r_{1}$, and
$x^{\prime}$ represents $t_{2}$ and $r_{2}$, since the coordinate $\phi$ is
taken as constant in the calculation of the correlation functions.
$\eta=+1,-1$ corresponding to the twisted and untwisted fields respectively,
and $\mathcal{N}=\sum_{p}\eta^{2p}$ is a normalization factor. In particular,
the field of BTZ black hole spacetime $\hat{\varphi}=\frac{1}{\sqrt
{\mathcal{N}}}\sum_{p}\eta^{p}\hat{\psi}\left(  \Gamma^{p}x\right)  $ is an
automorphic field that is constructed from a massless scalar field $\hat{\psi
}$ in (2+1) dimensional AdS spacetime via the identification $\Gamma
^{p}:\tilde{\phi}\rightarrow\tilde{\phi}+2\pi p\sqrt{M}$ \cite{foo2022quantum}%
. The function $\sigma(x,x^{\prime})$ is the squared geodesic distance between
$x$ and $x^{\prime}$, and is given in the BTZ black hole spacetime as
\begin{equation}
\sigma\left(  {\Gamma}_{1}^{p}x,{\Gamma}_{1}^{q}x^{\prime}\right)
=\frac{r_{1}r_{2}}{M\ell^{2}}\cosh\left[  2\pi(p-q)\sqrt{M}\right]
-1-\sqrt{\frac{r_{1}^{2}}{M\ell^{2}}-1}\sqrt{\frac{r_{2}^{2}}{M\ell^{2}}%
-1}\cosh\sqrt{M}\frac{t_{1}-t_{2}}{\ell}.
\end{equation}
What needs to be noted in the summation of the Wightman functions for the BTZ
black hole is the coincidence limit. When $x\rightarrow x^{\prime}$, only the
$p=q=0$ term in the sum will diverge, and this term can be subtracted to
renormalize the stress-energy tensor, effectively setting the energy of the
pure anti-de Sitter universe to zero \cite{carlip19952+}.

According to Ref. \cite{foo2022quantum}, we can extend the Wightman function
for the classical BTZ black hole to that for the mass-superposed BTZ black
hole, $W_{BTZ}^{(1,2)}\left(  x,x^{\prime}\right)  =\frac{1}{\mathcal{N}}%
\sum_{p,q}\eta^{p}\eta^{q}\langle0|\hat{\psi}\left(  \Gamma_{1}^{p}x\right)
\hat{\psi}\left(  \Gamma_{2}^{q}x^{\prime}\right)  |0\rangle$, which is given
explicitly as
\begin{equation}
W_{BTZ}^{(1,2)}\left(  x,x^{\prime}\right)  =\frac{1}{\mathcal{N}}\sum_{p}%
\sum_{q}\eta^{p}\eta^{q}\frac{1}{4\pi\ell\sqrt{2}}\left(  \frac{1}%
{\sqrt{\sigma\left(  {\Gamma}_{1}^{p}x,{\Gamma}_{2}^{q}x^{\prime}\right)  }%
}-\frac{\zeta}{\sqrt{\sigma\left(  {\Gamma}_{1}^{p}x,{\Gamma}_{2}^{q}%
x^{\prime}\right)  +2}}\right)  , \label{wf12}%
\end{equation}
where the cross term of squared geodesic distance is
\begin{align}
\sigma\left(  {\Gamma_{1}}^{p}x,{\Gamma_{2}}^{q}x^{\prime}\right)  =  &
\frac{r_{1}r_{2}}{\sqrt{M_{1}}\sqrt{M_{2}}\ell^{2}}\cosh\left[  2\pi
(p\sqrt{M_{1}}-q\sqrt{M_{2}})\right]  -1\nonumber\\
&  -\sqrt{\frac{r_{1}^{2}}{M_{1}\ell^{2}}-1}\sqrt{\frac{r_{2}^{2}}{M_{2}%
\ell^{2}}-1}\cosh\frac{\sqrt{M_{1}}t_{1}-\sqrt{M_{2}}t_{2}}{\ell},
\end{align}
where ${\Gamma_{1}}^{p}$ and ${\Gamma_{2}}^{q}$ are two different periodic
isometries with two different mass $M_{1}$ and $M_{2}$. This presents the
mass-superposed black hole, in which each mass defines a unique classical
solution to the Einstein field equations, and the amplitude obtained from the
mass superposition corresponds to the associated spacetime state
\cite{foo2022quantum}. In these scenarios, the horizon undergoes a
superposition characterized by $r_{s}=\sqrt{M_{1}}\ell$ and $r_{b}=\sqrt
{M_{2}}\ell$, defined by the two mass values $M_{1},M_{2}$ (assume
$M_{1}<M_{2}$ in our paper) of the black hole mass superposition. It is noted
that when the superposed mass takes the same value ($M_{1}=M_{2}=M$) in Eq.
(\ref{wf12}), the Wightman function reduces to the same form as in Eq.
(\ref{wf11}). This is not difficult to understand, because the two masses are
superposed at the same center point, as explained explicitly in Ref.
\cite{arabaci2022studying}. So, if these two masses are equal, the black hole
formed by their superposition in a certain proportion is equivalent to a
classical black hole of this mass. Moreover, the metric for the
mass-superposed black holes can be reconstructed through the Wightman function
\cite{saravani2016spacetime}, as discussed in detail in Ref.
\cite{foo2022quantum}.

In order to study the origin of Hawking radiations, we have to introduce some
correlation functions involving the derivative of Wightman functions or
two-point correlation functions, since the peak in these correlation functions
can represent the most possible creation position for Hawking particles from a
gravitational black hole. Here we introduce a correlation function as%
\begin{equation}
G_{\tau_{1}\tau_{2}}=\partial_{\tau_{1}}\partial_{\tau_{2}}\langle
0|\hat{\varphi}(x)\hat{\varphi}(x^{\prime})|0\rangle=\partial_{\tau_{1}%
}\partial_{\tau_{2}}W_{BTZ}\left(  x,x^{\prime}\right)  , \label{cfde}%
\end{equation}
where $W_{BTZ}\left(  x,x^{\prime}\right)  $ is the Wightman function
$W_{BTZ}^{(1,1)}\left(  x,x^{\prime}\right)  $ of a classical BTZ black hole
or the Wightman function $W_{BTZ}^{(1,2)}\left(  x,x^{\prime}\right)  $ of a
mass-superposed BTZ black hole. $G_{\tau_{1}\tau_{2}}$ is the velocity
correlation function, which can be used to describe the generation of
particles near the horizon \cite{anderson2022horizons}. Actually, one can also
define another correlation function as $G_{r_{1}r_{2}}=\partial_{r_{1}%
}\partial_{r_{2}}W_{BTZ}\left(  x,x^{\prime}\right)  $ based on the momentum
density, which describes the correlation between Hawking particles and
corresponding partner particles \cite{unruh1977origin}. For our purpose of
studying the creation of Hawking particles, $G_{\tau_{1}\tau_{2}}$ is proper.

For mass-superposed black holes, the position of the horizon is uncertain. So
we define the outside of the mass-superposed black hole as the outside beyond
the horizon given by the larger mass, that is the region $r_{2}>r_{b}%
=\sqrt{M_{2}}\ell$, and the inside means the region $r_{1}<r_{s}=\sqrt{M_{1}%
}\ell$. The region between $r_{s}$ and $r_{b}$ represents the uncertainty of
the horizon, which cannot be interpreted clearly in the present theory and may
require the complete quantum gravity theory. On the other hand, it has to be
interpreted how to define the time in the mass-superposed black hole
spacetime, in which the time is also uncertain. In our study, $\tau_{1}$
($\tau_{2}$) is linked to the spacetime of the BTZ black hole with the mass
$M_{1}$ ($M_{2}$), as implied in Eq. (\ref{wf12}). Thus, one can assign a
range of values to $\tau_{1}$ and another range of values to $\tau_{2}$ to
define their respective time. But all our calculations about the correlation
function are made under the condition of the equal time $\tau_{1}=\tau_{2}$.
This has been enough to give some meaningful results as presented in the third
and fourth section.

With these, it is possible to extend the theory in Ref. \cite{foo2022quantum}
which is calculated at one position where the Unruh-DeWitt detector is placed
to the two-point correlation function. When both points $r_{1}$ and $r_{2}$ of
the correlation function are outside the horizon and $r_{1}=r_{2}$, the
correlation peak can indicate the resonance effect for the mass-superposed
black hole spacetime, as presented in Ref. \cite{foo2022quantum} and the next
section of our paper. When one point of the correlation function is within the
horizon and the other point is outside the horizon, the correlation peak
indicates the correlation between Hawking particles and corresponding partner
particles for the mass-superposed black hole. This can be used to reveal the
creation position of Hawking particles.

Finally, the choice of observers is very important for calculating the
correlation. The stationary observer outside the BTZ black hole is a natural
choice by fixing the radial and angular coordinates to constants $R_{D}$ and
$\Phi_{D}$ as \cite{smith2014looking}
\begin{equation}
(t=\tau/\gamma_{D},r=R_{D},\phi=\Phi_{D}), \label{eq.so}%
\end{equation}
where $\gamma_{D}=\sqrt{\frac{R_{D}^{2}}{\ell^{2}}-M}$ is the red-shift
factor. We will reinvestigate the quantum signatures of the mass-superposed
BTZ black hole by a static observer outside this black hole in the next section.

Another important observer is the free-falling observer that freely falls into
the black hole, which plays an important role in studying the black hole
quantum atmosphere, since there is no static observer inside a gravitational
black hole. In the fourth section of this paper, we will consider a class of
free-falling observers falling from infinity and towards a singularity at
$r=0$. For this purpose, the Painlev\'{e}-Gullstrand (PG) coordinate system is
required, which is constructed based on the proper time of the free-falling
observer and is regular at the horizon \cite{gallock2021harvesting}.
Painlev\'{e}-type coordinates can be obtained by making the transformation
\begin{equation}
dt=d\tau-\frac{\sqrt{1-F}}{\Delta}dr, \label{tt}%
\end{equation}
where the function $F$ can be chosen as $F=\Delta=\frac{r^{2}}{\ell^{2}}-M$ in
the BTZ black hole, but it is not unique \cite{wu2006remarks}. Alternatively,
we can make a different choice of the function $F=\frac{\Delta}{\Delta_{0}}%
\,$with $\Delta_{0}=\Delta|_{M=0}=\frac{r^{2}}{\ell^{2}}$. With this choice
for the coordinate transformations (\ref{tt}), the line element (\ref{btzm})
becomes
\begin{equation}%
\begin{split}
ds^{2}  &  =-\Delta d\tau^{2}+2\sqrt{1-\frac{\Delta}{\Delta_{0}}}d\tau
dr+\frac{dr^{2}}{\Delta_{0}}+r^{2}d\phi^{2}\\
&  =-(\frac{r^{2}}{\ell^{2}}-M)d\tau^{2}+2\sqrt{\frac{M\ell^{2}}{r^{2}}}d\tau
dr+\frac{\ell^{2}}{r^{2}}dr^{2}+r^{2}d\phi^{2},
\end{split}
\label{cfo}%
\end{equation}
where
\begin{equation}
t=\tau+\frac{\ell}{\sqrt{M}}\ln(r)-\frac{\ell}{2\sqrt{M}}\ln\left\vert
r^{2}-M\ell^{2}\right\vert . \label{fo}%
\end{equation}
Note that $t=\tau+\frac{\ell}{\sqrt{M}}\ln(r)-\frac{\ell}{2\sqrt{M}}\ln
(r^{2}-M\ell^{2})$ represents the coordinates of the partner particle within
the horizon of the BTZ black hole, and $t=\tau+\frac{\ell}{\sqrt{M}}%
\ln(r)-\frac{\ell}{2\sqrt{M}}\ln(M\ell^{2}-r^{2})$ represents the coordinates
of the Hawking particle outside the horizon of the BTZ black hole.

\section{Quantum signatures}

In the last section, we have given the definition of the correlation function
which will be used in our paper, and discussed the role played by the
observers in the calculation of the correlation. In this section, we will use
the correlation functions and the stationary observers obtained in the last
section to revisit the quantum signatures of the black hole mass
superposition. We will investigate whether the earlier results can be
recovered using the correlation functions and whether there are some new
results that have not been revealed before.

\subsection{Obtaining quantum signals through velocity correlation function}

Through Eqs. (\ref{wf12}) and (\ref{cfde}), we can obtain the velocity
correlation function of the mass-superposed BTZ black hole as
\begin{equation}
G_{\tau_{1}\tau_{2}}=\partial_{\tau_{1}}\partial_{\tau_{2}}\frac
{1}{\mathcal{N}}\sum_{p}\sum_{q}\eta^{p}\eta^{q}\frac{1}{4\pi\ell\sqrt{2}%
}\left(  \frac{1}{\sqrt{\sigma\left(  {\Gamma}_{1}^{p}x,{\Gamma}_{2}%
^{q}x^{\prime}\right)  }}-\frac{\zeta}{\sqrt{\sigma\left(  {\Gamma}_{1}%
^{p}x,{\Gamma}_{2}^{q}x^{\prime}\right)  +2}}\right)  , \label{og}%
\end{equation}
where
\begin{align}
\sigma\left(  {\Gamma_{1}}^{p}x,{\Gamma_{2}}^{q}x^{\prime}\right)  =  &
\frac{r_{1}r_{2}}{\sqrt{M_{1}}\sqrt{M_{2}}\ell^{2}}\cosh\left[  2\pi
(p\sqrt{M_{1}}-q\sqrt{M_{2}})\right]  -1\nonumber\\
&  -\sqrt{\frac{r_{1}^{2}}{M_{1}\ell^{2}}-1}\sqrt{\frac{r_{2}^{2}}{M_{2}%
\ell^{2}}-1}\cosh\left(  \frac{\sqrt{M_{1}}\tau_{1}}{\ell\sqrt{\frac{r_{1}%
^{2}}{\ell^{2}}-M_{1}}}-\frac{\sqrt{M_{2}}\tau_{2}}{\ell\sqrt{\frac{r_{2}^{2}%
}{\ell^{2}}-M_{2}}}\right)  . \label{ogd}%
\end{align}
In order to facilitate comparison with Ref. \cite{foo2022quantum}, we use the
transparent boundary condition ($\zeta=0$) in the calculation of correlation.

Black hole mass superposition induces novel effects that are absent in
classical spacetime. As shown in Ref. \cite{foo2022quantum}, the detector
exhibits resonances at specific mass ratios when interacting with the Hawking
radiation from mass-superposed BTZ black holes. These resonances can be
understood in light of Bekenstein's conjecture regarding the quantization of
black hole mass \cite{bekenstein1973black}, in which the black hole event
horizon region and its mass are adiabatically invariant, with associated
discrete quantized and uniformly distributed energy levels. Here, we show that
quantum signatures can also be observed through the velocity correlation
function, as depicted in Fig. 1, which corroborates the results of Ref.
\cite{foo2022quantum} for the mass ratio $M_{1}/M_{2}=1/n^{2}$. Furthermore,
it is evident that the value of the equal-time velocity correlation peak
gradually decreases with increasing $n$ in the mass ratio. This indicates that
a larger difference between the two mass values in the black hole mass
superposition leads to reduced velocity correlation peaks. Because the larger
mass part will gradually dominate, resulting in a decrease in the velocity
correlation peak, which is similar to the reduction of the resonance phenomenon.

\begin{figure}[ptb]
\centering
{\includegraphics[width=0.68\textwidth]{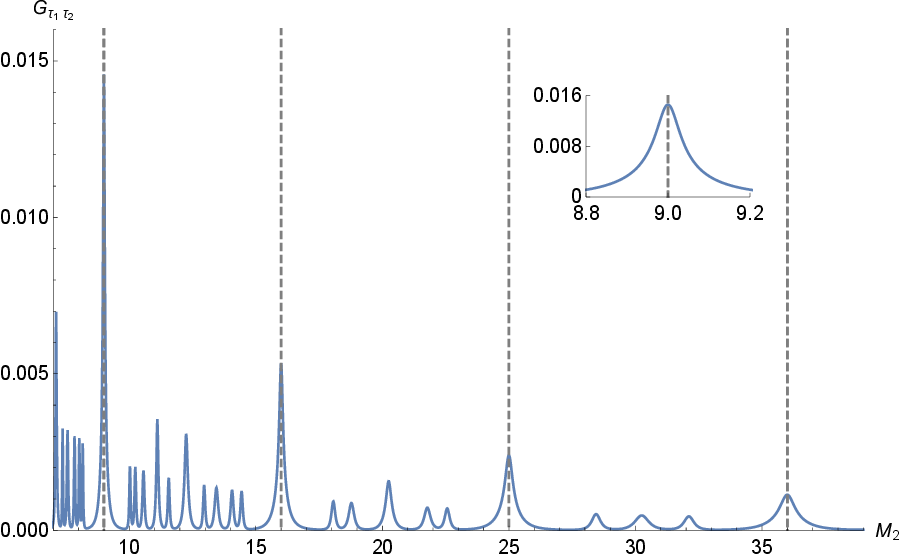}}\caption{The absolute value
of $G_{\tau_{1}\tau_{2}}$ as the function of $M_{2}$ with the mass ratio as
$M_{1}/M_{2}=1/n^{2}$ for the stationary observer outside the horizon of
mass-superposed black holes when the mass $M_{1}=1$ is fixed. The four high
peaks marked with the vertical dash lines correspond to the cases $n=3,4,5,6$,
respectively. In the upper right corner, we show a local magnified image with
a mass ratio of $(M_{1}=1, M_{2}=9)$. Other parameters are taken as
$r_{1}=r_{2}=10$, $\ell=1$, $p=q=20$ and $\tau_{1}=\tau_{2}=0$.}%
\label{fig1}%
\end{figure}

\begin{figure}[ptb]
\centering
{\includegraphics[width=0.68\textwidth]{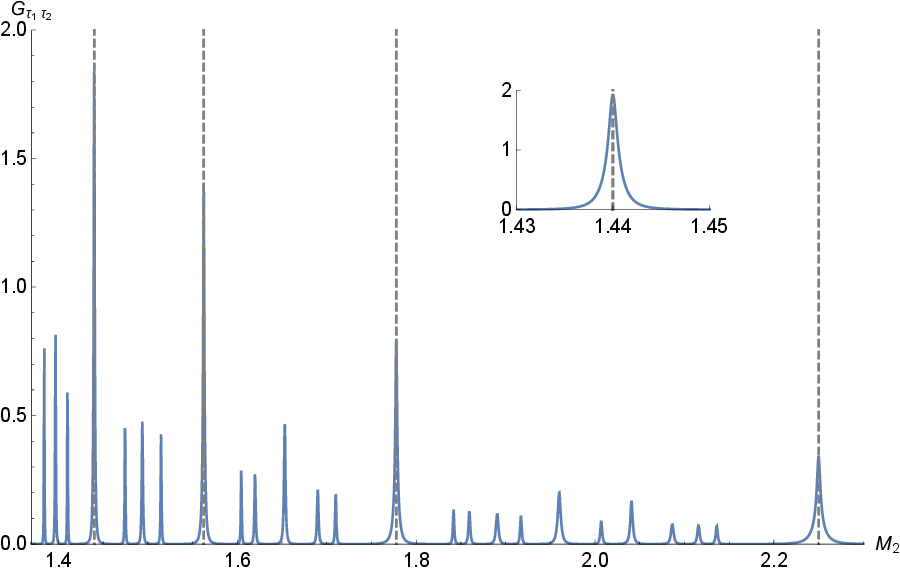}}\caption{The absolute value
of $G_{\tau_{1}\tau_{2}}$ as the function of $M_{2}$ with the mass ratio as
$M_{1}/M_{2}=(n-1)^{2}/n^{2}$ for the stationary observer outside the horizon
of mass-superposed black holes when the mass $M_{1}=1$ is fixed. The four high
peaks marked with the vertical dash lines correspond to the cases $n=3,4,5,6$,
respectively. In the upper right corner, we show a local magnified image with
a mass ratio of $(M_{1}=1, M_{2}=36/25)$. Other parameters are kept consistent
with those in Fig. 1.}%
\label{fig2}%
\end{figure}

Regarding the resonance phenomenon observed at the mass ratio $M_{1}%
/M_{2}=(n-1)^{2}/n^{2}$ as reported in Ref. \cite{foo2022quantum}, we
replicate these results through calculations of the equal-time velocity
correlation, illustrated in Fig. 2. In contrast to the scenario depicted in
Fig. 1 at the mass ratio $M_{1}/M_{2}=1/n^{2}$, we observe that the value of
the equal-time velocity correlation peak increases gradually as $n$ increases
in the mass ratio $M_{1}/M_{2}=(n-1)^{2}/n^{2}$. This is easy to understand,
because the two mass ratios have different trends with increasing $n$. In
particular, for each mass ratio, the correlation peak values increase as the
two mass values in the black hole mass superposition become closer to each other.

In Fig. 1 and Fig. 2, we use dashed lines to mark the quantum signals that
appear at special mass ratios. This indicates that the superposition spacetime
resonates at integer $\sqrt{M_{1}/M_{2}}$. The reason can be seen from Eq.
(\ref{og}) and (\ref{ogd}) where the term related to $cosh$ function
disappears at $p\sqrt{M_{1}}=q\sqrt{M_{2}}$. Such phenomena in quantum
correlations again support the scenario of a quantized BTZ black hole horizon
(without angular momentum) in \cite{6}, which could be gotten by the quantized
radius $r_{H}=\sqrt{M}\ell=n$ ($n=1,2,\ldots$).

In addition, we find that the equal-time velocity correlation also exhibits
peaks at certain mass ratios that lie between the special values $1/n^{2}$ and
$(n-1)^{2}/n^{2}$, but these peaks are smaller than those observed at these
specific mass ratios. Furthermore, the values of $p$ and $q$ in Eqs.
(\ref{og}) affect the occurrence of these correlation peaks. Larger values of
$p$ and $q$ result in more peaks in the velocity correlation for
mass-superposed black holes. When the radial coordinates $r_{1}=r_{2}$, the
equal-time velocity correlation peaks diminish as the radial coordinate moves
further away from the horizon $r_{b}=\sqrt{M_{2}}\ell$. In particular, the
correlation function will diverge at $M_{1}=M_{2}$ when $r_{1}=r_{2}$, but the
quantum signals of the mass-superposed black hole can still appear at these
different mass if the distance is not so far from the horizon $r_{b}$.

\subsection{Distinguishing quantum signals with same mass ratio}

It has been observed that resonant peaks occur at specific mass ratios, but
distinguishing between different mass pairs with the same mass ratio, such as
$(M_{1}=0.5,M_{2}=2)$, $(M_{1}=1,M_{2}=4)$ corresponding to $M_{1}%
/M_{2}=1/n^{2}$ with $n=2$, has not been discussed before. Here, we will use
the velocity correlation method to investigate it.

In Table 1, we compute and present four absolute values of $G_{\tau_{1}%
\tau_{2}}$ for superposed mass values: $(M_{1}=4,M_{2}=16)$, $(M_{1}%
=1,M_{2}=4)$, $(M_{1}=4,M_{2}=1)$, and $(M_{1}=0.5,M_{2}=2)$. From this table,
it is evident that different absolute values are obtained for the velocity
correlation even at the same mass ratios in the first, second, and fourth
rows. Furthermore, these absolute values of the correlation decrease as the
two mass values become larger for the same mass ratio, since larger mass
represents a stronger gravitational field which make it easier to reduce the
correlation from the black hole mass superposition. Moreover, it is noted from
the table that the absolute value of $G_{\tau_{1}\tau_{2}}$ for the mass
values $(M_{1}=1,M_{2}=4)$ is identical to that for $(M_{1}=4,M_{2}=1)$. This
indicates that these configurations are indistinguishable, demonstrating an
exchange symmetry in the velocity correlation functions for the
mass-superposed parameters $M_{1}$ and $M_{2}$. It also shows that our earlier
assumption, $M_{1}<M_{2}$, is reasonable and includes the phenomena appearing
for the case $M_{1}>M_{2}$. So, in the following calculation and discussion,
we will maintain this assumption without loss of generality.

\begin{table}[ptb]
\caption{The absolute value of $G_{\tau_{1}\tau_{2}}$ for the stationary
observer outside the horizon of mass-superposed black holes for four different
mass pairs: $(M_{1}=0.5, M_{2}=2)$, $(M_{1}=1, M_{2}=4)$, $(M_{1}=4,
M_{2}=16)$, and $(M_{1}=4, M_{2}=1)$. Other parameters are kept consistent
with those in Fig. 1. }%
\label{table1}%
\centering
\setlength{\tabcolsep}{6pt}
\begin{tabular}
[c]{ccccc}\hline
Mass ratios $(M_{1},M_{2})$ & $4, 16$ & $1, 4$ & $4, 1$ & $0.5, 2$\\\hline
$G_{\tau_{1}\tau_{2}} $ & $0.09753$ & $0.10875$ & $0.10875$ & $0.11064$%
\\\hline
\end{tabular}
\end{table}

It is noted that these quantum signals with the same mass ratio have only one
peak when measured by the Unruh-DeWitt detector \cite{foo2022quantum}, but
they have different numerical values given by the velocity correlation. Thus,
by comparing the velocity correlation values, it is helpful to deduce the
composition of two masses of the mass-superposed black hole.

\section{Creation and evolution of Hawking radiations}

The above analyses indicates that the mass ratio affects the velocity
correlation outside the horizon of mass-superposed black holes. In this
section, we will study the origin of Hawking radiation through velocity
correlations, just as researchers do when exploring the origin of Hawking
radiation in quantum atmospheres
\cite{parentani2010vacuum,balbinot2022quantum,anderson2022horizons,balbinot2023quantum,fontana2023stress}%
. To determine the origin of Hawking particles from the quantum atmosphere in
both classical BTZ black holes and mass-superposed BTZ black holes, it is
essential to examine the velocity correlation function across the horizon
\cite{anderson2022horizons}.

\subsection{Correlation across the horizon}

Consider the correlation between the Hawking particle outside the horizon and
its partner particle inside the horizon for the free-falling observer defined
by Eq. (\ref{cfo}) inside the event horizon. The inside and outside of the
horizon comply with our previous rules stated below Eq. (\ref{cfde}). Using
Eqs. (\ref{wf11}) and (\ref{cfde}), we get the velocity correlation function
of the classical BTZ black hole as
\begin{equation}
G_{\tau_{1}\tau_{2}}=\partial_{\tau_{1}}\partial_{\tau_{2}}\frac
{1}{\mathcal{N}}\sum_{p}\sum_{q}\eta^{p}\eta^{q}\frac{1}{4\pi\ell\sqrt{2}%
}\left(  \frac{1}{\sqrt{\sigma\left(  {\Gamma}_{1}^{p}x,{\Gamma}_{1}%
^{q}x^{\prime}\right)  }}-\frac{\zeta}{\sqrt{\sigma\left(  {\Gamma}_{1}%
^{p}x,{\Gamma}_{1}^{q}x^{\prime}\right)  +2}}\right)  , \label{vcfc}%
\end{equation}
where
\begin{align}
&  \sigma\left(  {\Gamma}_{1}^{p}x,{\Gamma}_{1}^{q}x^{\prime}\right)
=\frac{r_{1}r_{2}}{M\ell^{2}}\cosh\left[  2\pi(p-q)\sqrt{M}\right]
-1-\sqrt{\frac{r_{1}^{2}}{M\ell^{2}}-1}\sqrt{\frac{r_{2}^{2}}{M\ell^{2}}%
-1}\nonumber\\
&  \times\cosh\left\{  \left(  \frac{\sqrt{M}\tau_{1}}{\ell}+\ln(r_{1}%
)-\frac{1}{2}\ln(M\ell^{2}-r_{1}^{2})\right)  -\left(  \frac{\sqrt{M}\tau_{2}%
}{\ell}+\ln(r_{2})-\frac{1}{2}\ln(r_{2}^{2}-M\ell^{2})\right)  \right\}  .
\end{align}
Furthermore, using Eqs. (\ref{wf12}) and (\ref{cfde}), the velocity
correlation function of the mass-superposed BTZ black hole is gotten as
\begin{equation}
G_{\tau_{1}\tau_{2}}=\partial_{\tau_{1}}\partial_{\tau_{2}}\frac
{1}{\mathcal{N}}\sum_{p}\sum_{q}\eta^{p}\eta^{q}\frac{1}{4\pi\ell\sqrt{2}%
}\left(  \frac{1}{\sqrt{\sigma\left(  {\Gamma}_{1}^{p}x,{\Gamma}_{2}%
^{q}x^{\prime}\right)  }}-\frac{\zeta}{\sqrt{\sigma\left(  {\Gamma}_{1}%
^{p}x,{\Gamma}_{2}^{q}x^{\prime}\right)  +2}}\right)  , \label{vcfq}%
\end{equation}
where
\begin{align}
&  \sigma\left(  {\Gamma_{1}}^{p}x,{\Gamma_{2}}^{q}x^{\prime}\right)
=\frac{r_{1}r_{2}}{\sqrt{M_{1}}\sqrt{M_{2}}\ell^{2}}\cosh\left[  2\pi
(p\sqrt{M_{1}}-q\sqrt{M_{2}})\right]  -1-\sqrt{\frac{r_{1}^{2}}{M_{1}\ell^{2}%
}-1}\sqrt{\frac{r_{2}^{2}}{M_{2}\ell^{2}}-1}\nonumber\\
&  \times\cosh\left\{  \left(  \frac{\sqrt{M_{1}}\tau_{1}}{\ell}+\ln
(r_{1})-\frac{1}{2}\ln(M_{1}\ell^{2}-r_{1}^{2})\right)  -\left(  \frac
{\sqrt{M_{2}}\tau_{2}}{\ell}+\ln(r_{2})-\frac{1}{2}\ln(r_{2}^{2}-M_{2}\ell
^{2})\right)  \right\}  .
\end{align}
For the two different situations, the point $r_{1}$ remains inside the horizon
and the point $r_{2}$ remains outside the horizon for the following numerical calculation.

\begin{figure}[ptb]
\centering
\begin{subfigure}[h]{0.45\textwidth}
		{\includegraphics[width=\textwidth]{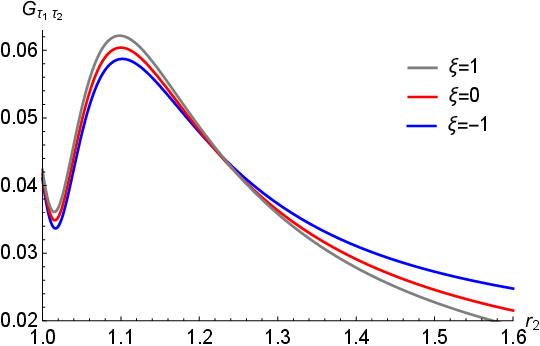}}
		\caption{}
		\label{4a}
	\end{subfigure}
\begin{subfigure}[h]{0.45\textwidth}
		{\includegraphics[width=\textwidth]{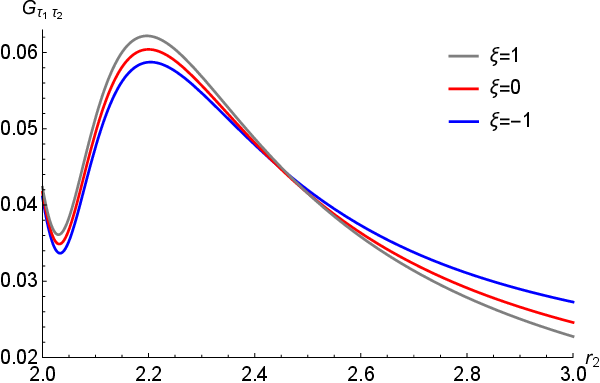}}
		\caption{}
		\label{4b}
	\end{subfigure}
\caption{The absolute value of $G_{\tau_{1}\tau_{2}}$ for the free-falling
observer crossing the horizon of the classical BTZ black hole (a) and the
mass-superposed BTZ black hole (b). The parameters are taken as $M=1$,
$r_{1}=0.9$ for the classical BTZ black hole (a) and $M_{1}=1, M_{2}=4$,
$r_{1}=0.9$ for the mass-superposed BTZ black hole (b). Neumann ($\xi=-1$),
transparent ($\xi=0$), and Dirichlet ($\xi=1$) boundary conditions are used
respectively. Other parameters are taken as $\ell=1$, $p=q=20$, and $\tau
_{1}=\tau_{2}=0$. }%
\label{fig3}%
\end{figure}

Using Eqs. (\ref{vcfc}) and (\ref{vcfq}), we depict the velocity correlation
function $G_{\tau_{1}\tau_{2}}$ of the free-falling observer passing through
the horizon of a classical BTZ black hole and a mass-superposition BTZ black
hole under three different boundary conditions, as presented in Fig. 3. It is
evident that a distinct peak appears near the event horizon $r_{h}=\sqrt
{M}\ell=1$ for the classical BTZ black hole, and near the horizon $r_{b}%
=\sqrt{M_{2}}\ell=2$ defined by the larger mass of the two mass values in the
black hole mass superposition. For both the classical and the mass-superposed
BTZ black holes, when $r_{2}$ approaches infinity, the value of the velocity
correlation function $G_{\tau_{1}\tau_{2}}$ will approach zero, but it is
finite at $r_{h}$ of the classical BTZ black hole or at $r_{b}$ of the
mass-superposed BTZ black hole. The appearance of the peak shows that the
quantum atmosphere can be investigated in the BTZ black hole and even the
mass-superposed BTZ black hole. That means that the location of the
correlation peak is just the location of the quantum atmosphere for the two
kinds of black holes.

We also present the different influence of different boundary conditions on
the peak value of the velocity correlation function in Fig. 3. It is noted
that the peak value is the largest when the boundary condition of the
correlation function is $\xi=1$, followed by $\xi=0$ and $\xi=-1$. But the
location of the peaks and the change trends are nearly the same for the three
boundary conditions, so in the following calculation we choose the boundary
condition $\xi=1$ which is enough for our purpose.

\begin{figure}[ptb]
\centering
\begin{subfigure}[h]{0.45\textwidth}
		{\includegraphics[width=\textwidth]{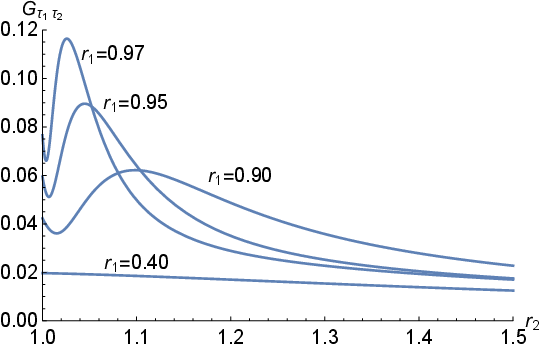}}
		\caption{}
		\label{4a}
	\end{subfigure}
\begin{subfigure}[h]{0.45\textwidth}
		{\includegraphics[width=\textwidth]{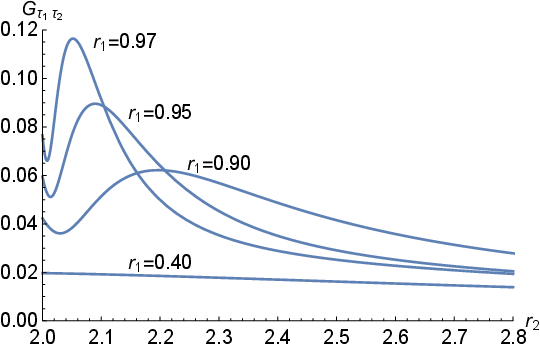}}
		\caption{}
		\label{4b}
	\end{subfigure}
\caption{The absolute value of $G_{\tau_{1}\tau_{2}}$ for the free-falling
observer crossing the horizon of the classical BTZ black hole (a) and the
mass-superposed BTZ black hole (b) for different values of $r_{1}$. The
parameters are taken as $M=1$ $(r_{1}=0.40,0.90,0.95,0.97)$ for the classical
BTZ black hole (a) and $M_{1}=1,M_{2}=4$ $(r_{1}=0.40,0.90,0.95,0.97)$ for the
mass-superposed BTZ black hole (b). Other parameters are taken as $\xi=1$,
$\ell=1$, $p=q=20$ and $\tau_{1}=\tau_{2}=0$. }%
\label{fig4}%
\end{figure}

At first, we investigate the change of the velocity correlation function with
$r_{2}$ for different values of $r_{1}$, as presented in Fig. 4. It is seen
that the velocity correlation peak increases when $r_{1}$ is close to the
horizon $r_{h}$ of the classical BTZ black hole from Fig. 4a, or when $r_{1}$
is close to the horizon $r_{s}=\sqrt{M_{1}}\ell$ defined by the smaller one of
the two masses in the mass superposition from Fig. 4b. When $r_{1}$ is far
away from the horizon $r_{h}$ of the classical BTZ black hole or far away from
the horizon $r_{s}$ of the mass-superposed BTZ black hole, it is found that
the velocity correlation peak disappears as shown in Fig. 4 ($r_{1}=0.4$).
This shows that the partner particles in classical BTZ black holes and
mass-superposed BTZ black holes will be swallowed up when they approach the
singularity as discussed in the background of classical Schwarzschild black
holes \cite{balbinot2022quantum}, so the correlation between the Hawking
particle and its partner particle disappears. Anyhow, the velocity correlation
function can reflect the position of the quantum atmosphere, even for
mass-superposed BTZ black holes.

\begin{figure}[ptb]
\centering
{\includegraphics[width=0.5\textwidth]{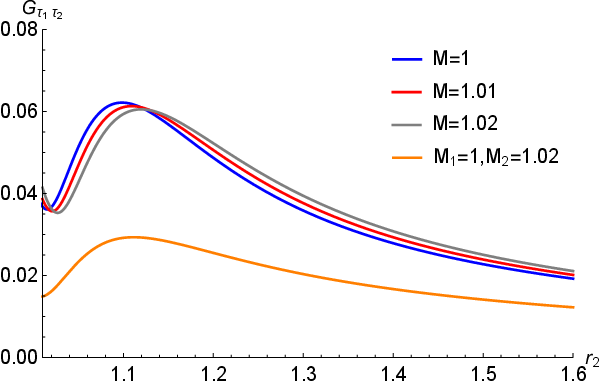}}\caption{The absolute value
of $G_{\tau_{1}\tau_{2}}$ for the free-falling observer crossing the horizon
of mass-superposed black holes $(M_{1}=1,M_{2}=1.02)$ and classical BTZ black
hole for three different mass values: $M=1$, $M=1.01$, and $M=1.02$. Other
parameters are taken as $\xi=1$, $r_{1}=0.9$, $\ell=1$, $p=q=20$, and
$\tau_{1}=\tau_{2}=0$.}%
\label{fig5}%
\end{figure}

\begin{figure}[ptb]
\centering
{\includegraphics[width=0.5\textwidth]{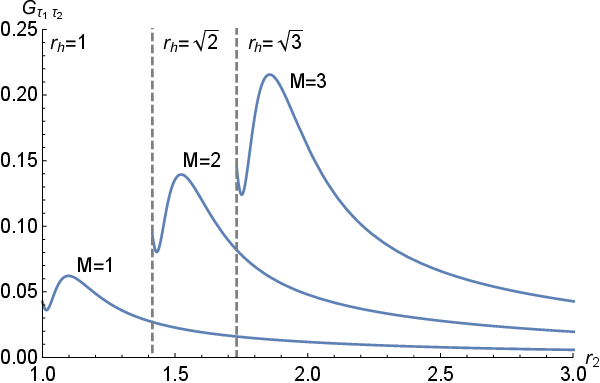}}\caption{The absolute value
of $G_{\tau_{1}\tau_{2}}$ for a free-falling observer crossing the event
horizon of the BTZ black hole of three different masses and different values
of $r_{1}$: $(M=1,r_{1}=0.9)$, $(M=2,r_{1}=1.3)$, and $(M=3,r_{1}=1.6)$. The
other parameters are $\xi=1$, $\ell=1$, $p=q=20$, and $\tau_{1}=\tau_{2}=0$.}%
\label{fig6}%
\end{figure}

Then, we considered the effect of the black hole mass on the velocity
correlation peak of the BTZ black holes. In order to compare them, we choose
three different classical BTZ black hole with masses $M=1,1.01,1.02$ and a
mass-superposed BTZ black hole with two masses $M_{1}=1$ and $M_{2}=1.02$.
Fig. \ref{fig5} presents the results, and it is seen that the correlation for
the mass-superposed black hole is smaller than that for all three classical
black holes. This is because the uncertainty of the horizon of the
mass-superposed black hole reduces the value of the correlation, since we do
not include the region between $r_{s}$ and $r_{b}$ in the calculation of the
correlation. Furthermore, it is found that the result from the mass
superposition is different from that from the average value ($\frac
{M_{1}+M_{2}}{2}$), as presented by the curves $M=1.01$ and that for the
mass-superposed black hole. This is consistent with the previous discussion
about the superposition of gravitational field made first by Penrose
\cite{rp1996}. At the same time, the calculation of correlation does not lead
to the collapse of the mass-superposed state, since the results for the
mass-superposed black hole are different from any one of the classical black
holes with $M=1,1.02$. Maybe, it is expected to compare the result of the
collapse for the mass-superposed black hole with that for the classical black
hole, but it is impossible in the present form since it is not known how to
collapse the mass superposition to some specific mass.

Moreover, it is seen from Fig. 5 that the peak values of the velocity
correlation for the three classical black holes decrease when the mass of the
black hole increases, which is not consistent with the change trend in the
Hawking temperature of the BTZ black hole, $\kappa=\frac{1}{2}\frac{\partial
f}{\partial r}_{r=r_{+}}=\frac{r_{+}}{\ell^{2}}$, $T=\frac{\kappa}{2\pi}%
=\frac{\sqrt{M}}{2\pi\ell}$. The reason is that the distance between the
horizon of the BTZ black hole and the fixed point $r_{1}=0.9$ increases when
the mass of the BTZ black hole increases, and the effect of the increase in
the distance is greater than the effect of the increase in the black hole mass
on the correlation peak. In order to make this clearer, we will calculate the
correlation for different combined values of $r_{1}$ and $M$.

When we increase the mass of the BTZ black hole and make $r_{1}$ close to the
horizon of the BTZ black hole, the velocity correlation peak will gradually
increase with the increase in mass as shown in Fig. 6, which indicates that
the amplitude of Hawking radiations generated by the quantum atmosphere of the
classical BTZ black hole will increase, which is consistent with the change
trend in the Hawking temperature of the BTZ black hole.

\begin{figure}[ptb]
\centering
{\includegraphics[width=0.5\textwidth]{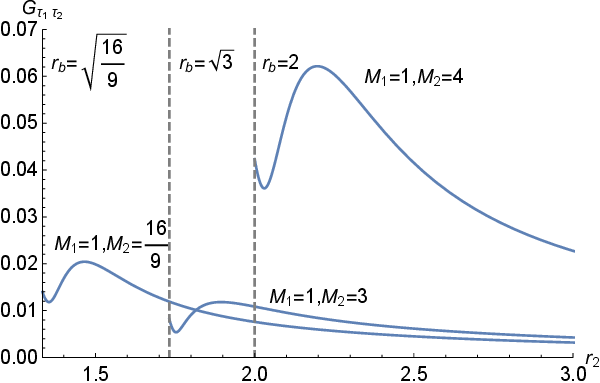}}\caption{The absolute value
of $G_{\tau_{1}\tau_{2}}$ for the free-falling observer crossing the horizon
of mass-superposed black holes for three different mass ratios: $(M_{1}%
=1,M_{2}=\frac{16}{9})$, $(M_{1}=1,M_{2}=3)$, and $(M_{1}=1,M_{2}=4)$. Other
parameters are taken as $\xi=1$, $r_{1}=0.9$, $\ell=1$, $p=q=20$, and
$\tau_{1}=\tau_{2}=0$.}%
\label{fig7}%
\end{figure}

For mass-superposed BTZ black holes, the mass ratio can play an important
role. It will not only affect the peak values, but also the peak position of
the velocity correlation. From Fig. 7, it is seen that when the mass ratio is
at special values such as $(M_{1}=1,M_{2}=4)$ and $(M_{1}=1,M_{2}=\frac{16}%
{9})$, the velocity correlation peak will be significantly larger than the
velocity correlation peak when the mass ratio is at the general value such as
$(M_{1}=1,M_{2}=3)$. This can be understood because the term related to $cosh$
function in Eq. (\ref{vcfq}) disappears at $p\sqrt{M_{1}}=q\sqrt{M_{2}}$. It
is also noted that in Fig. 7, the value of the correlation is larger for
larger $M_{2}$ with the same $M_{1}$, which is consistent with the change
trend for the classical black hole as presented in Fig. 6.

\subsection{Time evolution of correlation}

For classical BTZ black holes, the equal-time velocity correlation cannot
depict time evolution since the time term vanishes at equal times $\tau
_{1}=\tau_{2}$, as shown in Eq. (\ref{vcfc}). However, the time term remains
present in the velocity correlation for mass-superposed BTZ black holes, as
illustrated in Eq. (\ref{vcfq}). Therefore, we can analyze the time evolution
of Hawking radiation through the correlation function for mass-superposed BTZ
black holes.

\begin{figure}[ptb]
\centering
\begin{subfigure}[h]{0.45\textwidth}
		{\includegraphics[width=\textwidth]{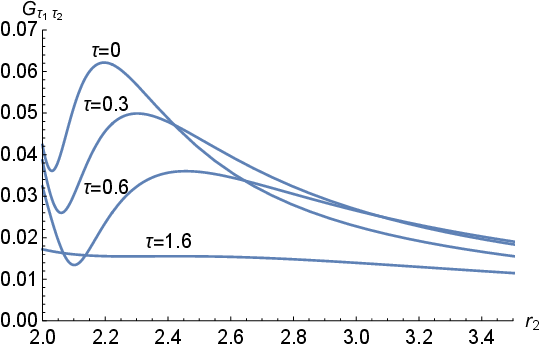}}
		\caption{}
		\label{8a}
	\end{subfigure}
\begin{subfigure}[h]{0.45\textwidth}
		{\includegraphics[width=\textwidth]{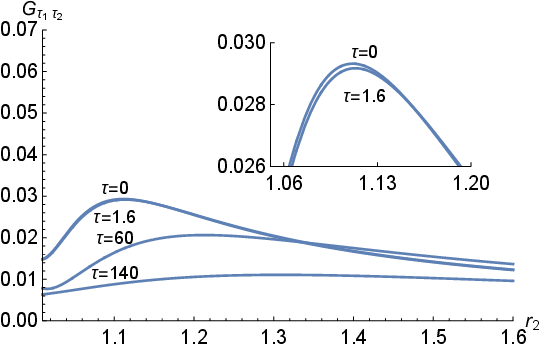}}
		\caption{}
		\label{8b}
	\end{subfigure}
\caption{The absolute value of $G_{\tau_{1}\tau_{2}}$ for the free-falling
observer crossing the horizon of mass-superposed BTZ black hole, the
parameters are taken as $M_{1}=1, M_{2}=4$, $r_{1}=0.9$, $(\tau_{1}=\tau
_{2}=0,0.3,0.6,1.6)$ (a) and the parameters are taken as $M_{1}=1, M_{2}%
=1.02$, $r_{1}=0.9$, $(\tau_{1}=\tau_{2}=0,1.6,60,140)$(b). A locally enlarged
image of $(\tau_{1}=\tau_{2}=0,1.6)$ is given in the upper right corner of
Figure (b). Other parameters are taken as $r_{1}=0.9$, $\xi=1$, $\ell=1$,
$p=q=20$. }%
\label{fig8}%
\end{figure}

In Fig. 8a, we present the absolute values of the velocity correlation at four
increasing times $\tau=0$, $\tau=0.3$, $\tau=0.6$, and $\tau=1.6$. For
correlations where one point is located inside $r_{s}$ and the other one is
located beyond $r_{b}$ defined by the larger mass of the two mass values, the
velocity correlation peak gradually decreases with time. Moreover, when
$\tau=1.6$, the correlation peak will nearly disappear. This shows that the
correlation of the mass-superposed black hole is lost, and the mass-superposed
black hole will no longer be in a superposition state. The decoherence or the
loss of correlation for the mass-superposed state of the black hole can be
understood to stem from the influence of Hawking radiations, as discussed in
Ref. \cite{aaz2019}. It is also noted that the velocity correlation peak
decreases very slowly with time when the two mass values in the mass
superposition are close,\ which means that the superposition state in this
case will be retained for a longer time. This is not hard to understand, since
a larger mass will dominate in the evolution if the values of two masses in
the mass superposition have large differences, which will be easier to lead to
the breakdown of the superposed state.

\section{Conclusions}

In this paper, we study the velocity correlation function in the context of
classical BTZ black holes and mass-superposed BTZ black holes. We recover the
quantum signatures of BTZ black hole mass superposition at specific mass
ratios, $\sqrt{M_{1}/M_{2}}=1/n$ and $\sqrt{M_{1}/M_{2}}=(n-1)/n$, by
observing the peaks in the equal-time velocity correlation function outside
the event horizon. These peaks diminish when the mass values in the black hole
mass superposition are significantly different. Notably, mass-superposed black
holes with the same mass ratio but different mass values can be distinguished
by the values of the correlation peaks, which has not been discussed before.

We also investigate the location and evolution of Hawking radiation for
mass-superposed BTZ black holes using the velocity correlation function. Since
the location of the peaks in the correlation function represents the location
where Hawking particle pairs are created, we find that the characteristic peak
appears near the horizon $r_{h}$ for classical BTZ black holes, and near
$r_{b}$ for mass-superposed BTZ black holes. As time progresses, the peaks in
the velocity correlation of the mass-superposed BTZ black holes gradually
decrease, which is similar to the behaviors of the decoherence of black hole
superpositions by Hawking radiations. In particular, when the values of the
two masses in the mass superposition are close, the correlation peak will
decrease very slowly with time and thus their superposition state will be
retained longer.

\section*{Acknowledgments}

This work is supported by National Natural Science Foundation of China (NSFC)
with Grant No. 12375057, and the Fundamental Research Funds for the Central
Universities, China University of Geosciences (Wuhan).

\end{document}